\documentclass[aps,superscriptaddress,nofootinbib]{revtex4}

\usepackage{graphicx}
\usepackage{dcolumn}

\newcommand{\half}{\frac{1}{2}}
\newcommand{\qq}{\langle \bar{q}q \rangle}

\newcommand{\qGq}{\langle \bar{q} g_s{\sigma}\cdot {G}q\rangle}

\newcommand{\GGb}{\langle  G^2 \rangle}

\def\aJLone<#1,#2>{#1}
\def\aJLtwo<#1,#2,#3>{#2}
\def\aJLyear<#1,#2,#3,#4>{#3}
\def\aJLpage<#1,#2,#3,#4>{#4}

\def\aJpage<#1,#2,#3>{#3}

\def\qbar{{\bar q}}
\def\g#1{\gamma^{#1}}
\def\Bt{{\tilde B}}

\begin{document}

\title{Chiral Symmetry of Nucleon Resonances in QCD sum rules}

\author{T.~Nakamura}%
\email[e-mail: ]{nakamura@th.phys.titech.ac.jp}%
\affiliation{%
Department of Physics, H-27, Tokyo Institute of Technology,\\ Meguro, Tokyo 152-8551, Japan
}%
\author{P.~Gubler}%
\email[e-mail; ]{gubler.p.aa@m.titech.ac.jp}%
\affiliation{Department of Physics, H-27, Tokyo Institute of Technology,\\ Meguro, Tokyo 152-8551, Japan
}%
\author{M.~Oka}%
\email[e-mail: ]{oka@th.phys.titech.ac.jp}%
\affiliation{%
Department of Physics, H-27, Tokyo Institute of Technology,\\ Meguro, Tokyo 152-8551, Japan
}%
\date{\today}

\begin{abstract}
The QCD sum rule approach is employed in order to study chiral properties of positive- and  negative-parity nucleon resonances.
It is pointed out that nucleons with an ``exotic'' chiral property, which can be represented by local five-quark operators,
can be paired with a standard nucleon forming a single chiral multiplet.
The sum rules of the five-quark operators, however,  are shown not to couple strongly to chirally-``exotic'' nucleon resonances
at the mass region of less than 2 GeV.
\end{abstract}
\maketitle

\section{introduction}
Chiral symmetry is one of the most important properties of low energy quantum chromodynamics (QCD).
It is an almost-exact symmetry of the QCD Lagrangian, although  not respected by the vacuum and the low-lying hadron spectrum. 
The QCD vacuum has a chirally-non-invariant condensate, $\qq$, and thus the symmetry is spontaneously broken\cite{NJL}.
The symmetry breaking is manifested as non-degenerate spectra of positive- and negative-parity hadrons, which would be identical if chiral symmetry were
exact and not broken. 
The symmetry breaking is accompanied by a Nambu-Goldstone boson, i.e., the pion, 
whose mass is not exactly zero, but small enough so that it is consistent with chiral symmetry.
Effects of chiral symmetry on the spectrum of the other mesons have been extensively studied\cite{HK}.
The scalar mesons are important ingredients  of chiral symmetry, which form a chiral multiplet together with the pseudoscalar mesons.
The vector mesons can be incorporated as hidden-gauge vector bosons, which obtain the masses by the Higgs mechanism\cite{KBY}.

In contrast, the baryons are not thoroughly studied in the context of chiral symmetry.
One way of treating the baryon consistently with chiral symmetry is to regard the baryon as a soliton state, ex. the Skyrmion\cite{Skyrmion,ANW}.
This picture is valid only for the large $N_c$ limit and generally $1/N_c$ corrections are not negligible for $N_c=3$.
Another formulation is an effective theory approach in which the baryon is classified by linear or non-linear representations of chiral symmetry\cite{Chiral_ET,JM}. 
It is the linear representation which is useful in studying the chiral properties of baryons.
In chiral SU(2)$\times$SU(2), the $I=1/2$ baryon (ex. nucleon) may belong to the fundamental representation, $(1/2,0)+(0,1/2)$, where the representation is given in terms of the isospin of the right SU(2) and that of the left SU(2).  This representation, called ``standard'', is identical to that of the fundamental quark.

A new formulation was proposed some time ago that involves two species of nucleons ($J=1/2$ and $I=1/2$)\cite{Lee,DeTar:1988kn,Jido}.
There one of them is standard, while the other one is nonstandard, belonging to the representation with the  left and right reversed, i.e., $(0,1/2) +(1/2,0)$.
In this ``exotic'' nucleon state, 
the {\it right}-handed component is transformed as a fundamental representation of the {\it left}-SU(2) and vice versa.
The combination of a standard nucleon and  an ``exotic'' nucleon leads to a possibility of paring positive- and negative-parity nucleons into  a single chiral representation (called chiral-mirror pair), 
so that in the limit of chiral symmetry restoration they form a degenerate nucleon pair.

The purpose of this paper is to study the possibility of the chiral-mirror pair of nucleons in the QCD sum rule technique, which 
has been proven useful in exploring hadron spectrum directly from QCD\cite{SVZ,RRY1985,Ioffe}.
Lattice QCD is obviously a powerful method in determining hadron masses from first principles, but it still has some limitations:
unbound resonances are not easily accessible, chiral symmetry is expensive, and unquenched calculations are not yet fully available.
The QCD sum rule is complementary in several ways, such that it is good at treating non-perturbative aspects of chiral symmetry breaking, and
that it gives direct relations between the hadron spectrum and the QCD vacuum properties.
We concentrate on the $N_f=2$ chiral symmetry and consider the chiral-mirror pair of nucleons, 
one of which is chirally-``exotic'', namely the left- and right-handed charges are reversely 
assigned compared with that of the fundamental quark.
We point out that such an assignment requires a five-quark interpolating field operator.
Construction of the sum rule for the five-quark operator is the main part of the work.

In sect.~II, we briefly review how the chiral-mirror pair can be classified in chiral SU(2)$_R\times$SU(2)$_L$.
In sect.~III, we consider the sum rule for the chirally-``exotic'' nucleon and point out that it can be represented by five-quark local operators. 
The sum rules are constructed for the five-quark operators also in sect.~III.
Results of the numerical analyses of the sum rules are presented in sect.~IV.  
Conclusions are given in sect.~V

\section{Representations for baryons}

Possible representations of the
chiral SU(2)$_R\times$SU(2)$_L$ symmetry for three quark baryons are 
given by\cite{Jido:1997yk,Cohen:1996sb}:
\begin{eqnarray}
&& q^3 \sim \left[q_r=\Bigl(\frac{1}{2},0\Bigr)\oplus 
q_l=\Bigl(0,\frac{1}{2}\Bigr)\right]^3
\nonumber \\
&&\hspace{1em}=
\left[\Bigl(\frac{3}{2},0\Bigr)\oplus\Bigl(0,\frac{3}{2}\Bigr)\right]\oplus
3\left[\Bigl(\frac{1}{2},1\Bigr)\oplus\Bigl(1,\frac{1}{2}\Bigr)\right]
\oplus 5\left[\Bigl(\frac{1}{2},0\Bigr)\oplus\Bigl(0,\frac{1}{2}\Bigr)\right]
\label{chi_QCD:18}
\end{eqnarray}
The first and third terms in r.h.s. correspond purely to isospin $3/2$
and $1/2$ states, respectively, while the second term gives 
a mixture of  isospin $1/2$ and $3/2$. 
The simplest choice for the nucleon is the 
$\Bigl(\frac{1}{2},0\Bigr)\oplus\Bigl(0,\frac{1}{2}\Bigr)$ representation.
Under the chiral group operation, this multiplet transforms in the same way as the fundamental quark field:
\begin{eqnarray}
i[Q_R^a,N_r]=-it^a\psi_r,\hspace{1em}i[Q_L^a,N_l]=-it^aN_l,
\label{chi_QCD:19}
\end{eqnarray}
where  $N_{r}$ and $N_{l}$ denote the right and left components of the iso-doublet nucleon field, $N=N_r+N_l$, respectively,
$Q_{R,L}^a$ ($a=1,2,3$) are the right- and left-SU(2) generators (charge operators) and 
$t^a \equiv \frac{\tau^a}{2}$ with $\tau^a$ being the Pauli matrices. 
The representations 
$\left[\Bigl(\frac{3}{2},0\Bigr)\oplus\Bigl(0,\frac{3}{2}\Bigr)\right]$ and 
$\left[\Bigl(\frac{1}{2},1\Bigr)\oplus\Bigl(1,\frac{1}{2}\Bigr)\right]$ are candidates for the $\Delta$
baryon resonance as both of them contain isospin $I=\frac{3}{2}$ states.
In a previous study\cite{Jido:1999hd}, the representation 
$\left[\Bigl(\frac{1}{2},1\Bigr)\oplus\Bigl(1,\frac{1}{2}\Bigr)\right]$
is chosen for $\Delta$.
This choice is consistent with $\Delta$ being a resonance strongly coupled to
the $N-\pi$ system, because 
$N\otimes\pi\sim\left[\Bigl(\frac{1}{2},0\Bigr)\oplus\Bigl(0,\frac{1}{2}\Bigr)\right]
\otimes\left[\Bigl(\frac{1}{2},\frac{1}{2}\Bigr)\right]$ does not contain 
$\left[\Bigl(\frac{3}{2},0\Bigr)\oplus\Bigl(0,\frac{3}{2}\Bigr)\right]$.
The physical baryons may well be mixed states of multiple representations, because
chiral symmetry is both explicitly and spontaneously broken\cite{Weinberg:1969hw,Gerstein}.
We, however, concentrate on the simplest representation, 
$\Bigl(\frac{1}{2},0\Bigr)\oplus\Bigl(0,\frac{1}{2}\Bigr)$ for the nucleon here, assuming 
that this is the main component.

Now, let us consider two nucleon species, $N_1$ and $N_2$, 
later being assigned to positive and negative parity nucleon resonances.
There are two
possible ways to assign the chiral representations for $N_1$ and $N_2$.
In the first case (called naive assignment\cite{Jido:1998av}),
both $N_1$ and $N_2$ belong to the same representation as that of
Eq. (\ref{chi_QCD:19}), i.e,
\begin{eqnarray}
&&i[Q_R^a,N_{1r}]=-it^aN_{1r},\hspace{1em}i[Q_L^a,N_{1l}]=-it^aN_{1l},
\nonumber\\
&&i[Q_R^a,N_{2r}]=-it^aN_{2r},\hspace{1em}i[Q_L^a,N_{2l}]=-it^aN_{2l}. \quad \hbox{(standard)}
\label{chi_QCD:22}
\end{eqnarray}
The second choice, called {\it mirror} assignment\cite{Jido:1998av}, is to give $N_2$ reversed charges as
\begin{eqnarray}
&&i[Q_R^a,N_{1r}]=-it^aN_{1r},\hspace{1em}i[Q_L^a,N_{1l}]=-it^aN_{1l},
\nonumber\\
&&i[Q_L^a,N_{2r}]=-it^aN_{2r},\hspace{1em}i[Q_R^a,N_{2l}]=-it^aN_{2l}. \quad \hbox{(exotic)}
\label{chi_QCD:23}
\end{eqnarray}
Note that the right-handed $N_{2r}$ transforms under $SU(2)_L$, while $N_{2l}$
transforms under $SU(2)_R$. 
We call the transformation rules in Eq. (\ref{chi_QCD:22}) and in Eq. (\ref{chi_QCD:23})
``standard" and ``exotic", respectively.
In the four component representation, 
we note that the axial-vector coupling constant $g_A$ of $N_2$ has the opposite sign to that of $N_1$
in the mirror assignment, i.e.,  
\begin{eqnarray}
&&
i[Q_A^a,N_1]=-i t^a\gamma_5 N_1, \qquad 
i[Q_A^a,N_2]=+i t^a\gamma_5 N_2.
\label{chi_QCD:25}
\end{eqnarray}

The possibility of the mirror assignment was first pointed out by Lee\cite{Lee}.
Later, DeTar and Kunihiro\cite{DeTar:1988kn} showed that the mirror model for the ground-state nucleon and 
its negative-parity excited state gives interesting features using the linear sigma model
formulation. 
%
%
The linear sigma model Lagrangian for the mirror assignment is given by
\begin{eqnarray}
\mathcal{L}_{{\rm{mirror}}}&=&\bar{N}_1i\partial\!\!\!/N_1
-g_1\bar{N}_1(\sigma+i\gamma_5\vec{\tau}\!\cdot\!\vec{\pi})N_1
\nonumber \\
&+&\bar{N}_2i\partial\!\!\!/N_2
-g_2\bar{N}_2(\sigma-i\gamma_5\vec{\tau}\!\cdot\!\vec{\pi})N_2
\nonumber \\
&-&m_0(\bar{N}_1\gamma_5N_2-\bar{N}_2\gamma_5N_1),
\label{Linear:5}
\end{eqnarray}
It should be noted that the last term of Eq.~(\ref{Linear:5}) gives a new mass term, which does not break chiral symmetry.
The mass parameter $m_0$ is a chiral scalar, whose origin in QCD is
very interesting.  It should not be related to the quark condensate, $\qq$, which breaks chiral symmetry, 
but should be given by other mass scales, such as the gluon condensate $\GGb$.

When chiral symmetry is spontaneously broken by the finite sigma condensate, $\langle\sigma\rangle=\sigma_0\neq0$,
the mass matrix of the Lagrangian in Eq. (\ref{Linear:5}) can be diagonalized, giving the mixing angle of $N_1$
and $N_2$ by 
\begin{eqnarray}
&&\tan 2\theta =\frac{2m_0}{\sigma_0(g_1+g_2)},\\
&&m_\pm =m_0 \sqrt{1+ \frac{\sigma_0^2}{4m_0^2} (g_1+g_2)^2}
\pm \frac{\sigma_0}{2} (g_1-g_2).
\label{Linear:6}
\end{eqnarray}
One sees from Eq.(\ref{Linear:6}) that the nucleons $N_+$
and $N_-$ have a finite mass $m_0$ when chiral symmetry is restored $\sigma_0=0$.
In contrast, for usual (Dirac) fermions chiral symmetry in the Wigner-Weyl mode
restricts particles to be massless, hence they acquire their masses only via the coupling
with the quark condensate of the vacuum. 

An interesting feature of the mirror assignment is the opposite signs of the axial-vector couplings of $N_+$ and $N_-$. 
As they are related to the pion-nucleon Yukawa coupling constants 
through the Goldberger-Treiman relation, they may be observed directly in pion-eta photo-production experiments\cite{Jido2}.
After the mixings of $N_1$ and $N_2$, this feature still remains as we see from the commutation relations,
\begin{eqnarray}
\left[Q_A^a,N_+\right]&=&t^a\gamma_5(\cos 2\theta N_+-\sin 2\theta \gamma_5N_-),
\label{Linear:7}\\
\left[Q_A^a,N_-\right]&=&t^a\gamma_5(-\sin 2\theta N_+-\cos 2\theta \gamma_5N_-),
\label{Linear:8}
\end{eqnarray}
giving the matrix expression
\begin{eqnarray}
g_A=\left( 
\begin{array}{cc}
\cos 2\theta & -\sin 2\theta \\
-\sin 2\theta  & -\cos 2\theta  
\end{array} 
\right).
\label{Linear:9}
\end{eqnarray}
From this we see the diagonal components $g_A^{++}$ and $g_A^{--}$ have opposite
signs as expected. The absolute values are smaller than unity. 

It is interesting to see how the mirror model works phenomenologically.
If we assign the ground state $N(939)$ and the first negative-parity excited state $N(1535)$ 
for $N_+$ and $N_-$, respectively, then
we can fix the model parameters, using $m_+=939$, $m_-=1535{\rm{MeV}}$,
$\sigma_0=f_\pi=93{\rm{MeV}}$ and $g_{\pi N_+ N_-}\sim 0.7$. 
The last relation is extracted
from the partial decay width of $N(1535)$\cite{Yao:2006px}.
This choice of parameters results in the mixing angle
$\theta=6.3^\circ$, and  the axial charges,
$g_A^{++}=-g_A^{--}=0.98$.
This  contradicts with experiment for the nucleon, $g_A\sim1.25$. 
It is, however, well known that the experimental value can be reproduced in the linear sigma model
only when higher dimensional representations, such as $(1,1/2)$, are introduced and
superposed to the $(1/2,0)+(0,1/2)$ representation\cite{Weinberg:1969hw}.
Hence the mixing of the two states $N_{1,2}$ is not very large;
the nucleon $N(939)$ is dominated in this scheme by the standard chiral component,
$N_1$, while $N(1535)$ by the ``exotic'' component, $N_2$.

\section{Analyses of ``exotic" nucleons in QCD sum rules}\label{chisum}

It is extremely interesting to check in QCD whether the mirror assignment is realized in the nucleon spectrum.
If it is the case, we expect to find a nucleon resonance state which couples to the local operator 
transformed as in Eq. (\ref{chi_QCD:23}).
In order to investigate the existence of ``exotic" nucleon states in the QCD sum rule technique, we start from constructing
local interpolating field operators with the relevant chiral properties, and then the sum rules are constructed and studied.

\subsection{Interpolating fields for  ``exotic" nucleons}

According to the PCAC relation, the pion corresponds to a local quark bilinear operator,
\begin{eqnarray}
\phi_{\pi}^{a}(x) \sim
\partial_{\mu}[\qbar(x) \g{\mu} \gamma_5 \tau^{a} q(x)]_{\rm{CS}}
\sim 2m_{q}[\qbar(x) i\gamma_{5} \tau^{a} q(x)]_{\rm{CS}} \, .
\end{eqnarray}
Here $[\qbar\ldots q]_{\rm{CS}}$ stands for the color-singlet $(\qbar\ldots q)$.
It is easy to see that this operator is transformed as the $a$-th
component of a chiral $O(4)$ vector.  The axial transformation gives
\begin{eqnarray}
    \null[Q_A^a, [\qbar i\gamma_{5}
    \tau^b q]_{\rm{CS}}] = -i \delta^{ab}
    [\qbar q]_{\rm{CS}}\nonumber\\
    \null[Q_A^a, [\qbar q]_{\rm{CS}}] = i [\qbar i\gamma_{5}
    \tau^a q]_{\rm{CS}}
    \label{eq:SPcomm}
\end{eqnarray}
Here $[\qbar q]_{\rm{CS}}$ is the fourth component of the $O(4)$
vector,  which corresponds to a scalar isoscalar meson, i.e. $\sigma$.
Thus the set of operators,
\begin{equation}
    \left( [\qbar(x) i\gamma_{5} \tau^{a} q(x)]_{\rm{CS}}\, ,
    \; [\qbar(x) q(x)]_{\rm{CS}} \right)
    \sim (\pi^a \, , \; \sigma)\, ,
    \label{eq:SPdef}
\end{equation}
belongs to a linear representation $(\half, \half)$ of the chiral
$SU(2)_{R} \otimes SU(2)_{L}$ group.

The nucleon operator is supposed to consist of
three quark operators:
\begin{equation}
    B^{\alpha} (x) \equiv \left[ (q^T(x) C \gamma_{5}\,q(x))_{I=0} q(x)^{\alpha}
    \right]_{\rm{CS}}
    \label{eq:B}
\end{equation}
where $\alpha$ is the Dirac index of the nucleon and $C$ is the charge
conjugation gamma matrix, $C= i \gamma^2\gamma^0$.
The first two quarks in (\ref{eq:B}) form a scalar diquark with $I=0$,
which commutes with $Q_A^{a}$.  Thus the
nucleon operator itself behaves just like a quark field under chiral
transform,
\begin{equation}
    [Q_A^a, B(x)] = - \gamma_{5} t^a B(x)
    \label{eq:QB}
\end{equation}
This equation proves that $B$ is chirally-standard so that it belongs to the linear representation,
$(\half,0)\oplus(0,\half)$, with the axial charge $g_A=1$.

There is an alternative choice for the nucleon operator,
\begin{equation}
    \Bt^{\alpha} (x) \equiv
    \left[ (q^T(x) C q(x))_{I=0} \{\gamma_{5}\,q(x)\}^{\alpha}
    \right]_{\rm{CS}} \, .
    \label{eq:BT}
\end{equation}
It is easy to see that $\Bt$ also satisfies Eq.(\ref{eq:QB}), and
is standard.
The most general local operator with no derivative (for the proton)
is given explicitly by
\begin{eqnarray}
B_p(x) = 
\epsilon^{abc} \left[ (u^{aT} C d^b)\,\gamma_{5} u^c
     + t (u^{aT} C\gamma_{5}\,d^b) \,u^c \right]
      ,
\label{bpgeneral}      
\end{eqnarray}
where $a$, $b$, and $c$ stand for the color of the quarks, and $t$
is a parameter which controls the mixing of the two independent
operators.

What is a possible local operator of quarks that transforms
in the chirally-``exotic'' way?
The signature of the chirally-``exotic'' nucleon is a
negative axial charge, $g_{A}<0$.
We have seen that the three-quark nucleon operators,
(\ref{bpgeneral}), give $g_A=1$, corresponding to the standard chiral representation.
It turns out that the ``exotic'' nucleon requires more than three quarks
if it is constructed as a local operator without derivatives.

We consider the color-singlet scalar/pseudoscalar bilinear operators,
given by (\ref{eq:SPdef}), which are transformed as (\ref{eq:SPcomm}).
Then, the ``exotic" nucleon can be represented by a local five-quark operator,
\begin{equation}
    B^{*\alpha} = [\qbar q ]_{\rm{CS}} B^{\alpha}
    + [\qbar i\gamma_{5} \tau^b q]_{\rm{CS}}
    (i \gamma_{5} \tau^b B)^{\alpha} \, ,
    \label{eq:Bmirror}
\end{equation}
where $B$ is the iso-doublet three-quark operators defined by Eq.~(\ref{eq:B}).
Using Eqs. (\ref{eq:QB}) and (\ref{eq:SPcomm}),
one easily confirms
\begin{equation}
    [Q_A^a, B^{*}] = +\gamma_5 t^a B^* \, ,
    \label{ga-1}
\end{equation}
and thus $B^*$ has a negative axial charge $g_A = -1$.
In constructing the sum rule, we employ $B^{*}$ for generating an ``exotic" nucleon state,
although the five-quark local operator for the ``exotic'' nucleon is not unique.

\subsection{QCD sum rules for ``exotic" nucleons}\label{SR_exo}
In constructing QCD sum rules for ``exotic" nucleons,
we do not assume parity of the nucleon so that both the positive and negative-parity
states with five-quarks can contribute to the sum rule.
If we obtain a negative-parity ``exotic'' nucleon, then it is considered to be paired with a 
positive-parity standard nucleon, while a positive-parity ``exotic'' nucleon should couple to
a negative-parity standard nucleon state.  Thus determining the parity of the obtained state
is important. 

It has been shown that the parity projection can be successfully performed for
the nucleon states using the retarded Green's function in the sum rule for the rest frame, $\vec{q}=0$,
\cite{Jido:1996ia}
\begin{eqnarray}
\Pi(q_0)=\int{\rm{d}}^4x\;e^{iq_0x^0}i\langle0|\theta(x^0)B^\ast(x)\bar{B}^\ast(0)|0\rangle.
\label{SR_exo:5}
\end{eqnarray}
This correlation function is analytic for Im $q_0>0$ and its imaginary part  for real $q_0>0$ can be expressed in terms of the
spectral functions as
\begin{eqnarray}
\frac{1}{\pi}\textrm{Im}\Pi(q_0)&=& \frac{\gamma_0+1}{2} \rho^+(q_0) +
\frac{\gamma_0-1}{2}   \rho^-(q_0) 
\nonumber\\
&=& A(q_0)\gamma_0+B(q_0),
\label{SR_exo:7}
\end{eqnarray}
where $\rho^+$ ($\rho^-$) is the spectral function for the positive- (negative-) parity nucleon 
states at $q_0>0$, and
\begin{eqnarray}
A(q_0)=\frac{1}{2}\left(\rho^+(q_0)+\rho^-(q_0)\right),\hspace{2em}
B(q_0)=\frac{1}{2}\left(\rho^+(q_0)-\rho^-(q_0)\right),
\label{SR_exo:8}
\end{eqnarray}
or equivalently,
$\rho^{\pm}(q_0)=A(q_0)\pm B(q_0)$.
It is important to note that the $B$ term is responsible for every difference in the positive- and negative-parity
spectral functions.  In the following, we observe in the operator product expansion (OPE) that 
the $B$ term is proportional to (odd powers of) 
chiral odd condensates, such as $\qq$ or $\qGq$.

The imaginary part of the correlation function is evaluated at the asymptotic region,
$q_0^2\to-\infty$ (or $q_0 \to  i\infty$) using the operator product expansion (OPE) technique. 
We obtain
\begin{eqnarray}
A_{{\rm{OPE}}}(q_0)&=&(90+36t+90t^2)\frac{q_0^{11}}{2^{11}\times 5!\;7!\pi^8}
\nonumber \\
&&+(41+14t+41t^2)\frac{q_0^7}{2^{10}\times 4!\;5!\pi^6}\langle\alpha_s\pi^{-1}G^2\rangle
\nonumber \\
&&+(111+42t+153t^2)\frac{q_0^5}{3\times 2^{12}\pi^4}\langle\bar{q}q\rangle^2
\nonumber \\
&&-(-25+38t+203t^2)\frac{q_0^3}{3\!\times\!2^{12}\pi^4}\langle\bar{q}q\rangle\langle\bar{q}g_s
\vec{\sigma}\!\cdot\!\vec{G}q\rangle
\nonumber \\
&&+(523+166t+1219t^2)\frac{q_0}{3^2\!\times\!2^{12}\pi^2}\langle\bar{q}q\rangle^2\langle\alpha_s\pi^{-1}G^2\rangle   
\nonumber \\
&&+(-3539+1150t+6853t^2)\frac{q_0}{3^2\!\times\!2^{17}\pi^4}\langle\bar{q}g_s\vec{\sigma}\!\cdot\!\vec{G}
q\rangle^2
\nonumber \\
&&-(131+50t-181t^2)\frac{q_0}{3^2\!\times\!2^4}\langle\bar{q}q\rangle^4\delta(q_0^2),
\nonumber \\
B_{{\rm{OPE}}}(q_0)&=&(-33-6t+39t^2)\frac{q_0^8}{2^{10}\times 4!\;5!\pi^6}\langle\bar{q}q\rangle
\nonumber \\
&&-(-13-7t+20t^2)\frac{q_0^6}{2^{15}\times 3^2\pi^6}\langle\bar{q}g_s\vec{\sigma}\!\cdot\!\vec{G}q\rangle
\nonumber \\
&&+(-28+5t+23t^2)\frac{q_0^4}{3^2\!\times\!2^{12}\pi^4}\langle\bar{q}q\rangle\langle\alpha_s\pi^{-1}G^2\rangle
\nonumber \\
&&-(-37-14t+45t^2)\frac{q_0^2}{2^7\pi^2}\langle\bar{q}q\rangle^3
\nonumber \\
&&+(-289+14t+119t^2)\frac{1}{3\!\times\!2^9\pi^2}\langle\bar{q}q\rangle^2\langle\bar{q}g_s\vec{\sigma}
\!\cdot\!\vec{G}q\rangle
\label{SR_exo:10}
\end{eqnarray}
from the OPE up to the dimension 12 operators. We take the chiral limit for the up and down
quarks, i.e. $m_q=0$, where we use the symbol $q$ for the up and down quarks.

A technical remark is made here on the QCD sum rule of five-quark interpolating operators.
In determining the validity of the constructed QCD sum rules, it is crucial to check the dominance of the pole contribution 
to the continuum contribution in the spectral function. 
In the case of five-quark sum rules, the continuum contributions are potentially large, since the logarithmic
terms appear in higher-order terms in the OPE. This stems from the higher mass dimension of
the correlation function than the ordinary nucleon case owing to the larger number of quark fields\cite{Nakamura}. 
To avoid this problem, the OPE should be calculated up to higher dimensional terms. 
In the analysis of the pentaquark sum rules, the importance of
the higher dimensional contributions of the OPE is examined in detail\cite{Kojo:2006bh}.

The sum rule is obtained by comparing the OPE of the correlation function, Eq. (\ref{SR_exo:10}),
and the explicit forms of the spectral functions at real positive $q_0$ via the analytic continuation.
It is assumed that
the spectral function can be expressed by a pole plus continuum contribution,
\begin{eqnarray}
\rho^\pm_{{\rm{phen}}}(q_0)=|\lambda^\pm|^2\delta(q_0-m^\pm)+\theta(q_0-{\sqrt{s_{{\rm{th}}}}})
\rho^\pm_{{\rm{cont}}}(q_0),
\label{SR_exo:11}
\end{eqnarray}
where $|\lambda^\pm|^2$ denotes the residue of the pole. The residue should be positive, 
which gives a condition to check the validity of the results obtained from the sum rule.
The continuum part is further assumed to be identical to the corresponding OPE function at above
the threshold $\sqrt{s_{{\rm{th}}}}$, $\rho_{{\rm{cont}}}^\pm=\rho_{{\rm{OPE}}}^\pm\equiv
A_{{\rm{OPE}}}\pm B_{{\rm{OPE}}}$.

In order to enhance the pole part and also suppress the higher dimensional
terms of the OPE, we introduce a weight function, which is analogous to the Borel transformation, 
a popular technique in QCD sum rules, 
as
\begin{eqnarray}
W(q_0)=\exp\left(-\frac{q_0^2}{M_B^2}\right).
\label{SR_exo:12}
\end{eqnarray}
$M_B$ is the relevant mass scale, which we call the Borel mass,
and the sum rule is obtained from
\begin{eqnarray}
\int {\rm{d}}q_0 W(q_0)\rho^\pm_{{\rm{phen}}}(q_0)=\int{\rm{d}}q_0W(q_0)\rho^\pm_{{\rm{OPE}}}
(q_0).
\label{SR_exo:13}
\end{eqnarray}  

Physical quantities are to be independent of the choice of $M_B$ ideally, but in practice,
the truncation in the OPE and the incompleteness of the pole plus continuum assumption
lead to mild dependence. We have to choose a reasonable range of $M_B$ carefully
to evaluate the physical quantities.
Finally, we obtain the sum rules for the positive- and negative-parity nucleons,
\begin{eqnarray}
L^+(M_B^2; s^+_{\rm{th}})=|\lambda^+|^2\exp{\left[-\frac{m_+^2}{M_B^2}\right]}=\int_0^{\sqrt{s^+_{{\rm{th}}}}}{\rm{d}}q_0\;
\rho^+_{\rm{OPE}}(q_0)\;\exp{\left[-\frac{q_0^2}{M_B^2}\right]}
\label{SR_exo:14} \\
L^-(M_B^2; s^-_{\rm{th}})=|\lambda^-|^2\exp{\left[-\frac{m_-^2}{M_B^2}\right]}=\int_0^{\sqrt{s^-_{{\rm{th}}}}}{\rm{d}}q_0\;
\rho^-_{\rm{OPE}}(q_0)\;\exp{\left[-\frac{q_0^2}{M_B^2}\right]}
\label{SR_exo:15}
\end{eqnarray}
The masses of the positive- and negative-parity states are obtained by taking the logarithmic derivative
of Eqs. (\ref{SR_exo:14}) and (\ref{SR_exo:15}):
\begin{eqnarray}
m^2_\pm(M_B^2; s_{\rm{th}})={\rm{d}}\log L^\pm/{\rm{d}}(-1/M_B^2).
\label{SR_exo:16}
\end{eqnarray} 
\section{Results}
\subsection{The sum rules for $B^*$}

The OPE of the correlator s expressed in terms of several parameters in QCD, such as the quark masses and the vacuum
condensates.
We take the chiral limit for the up and down quarks, i.e. $m_q=0$, for simplicity, 
because the finite light-quark masses (except strangeness) hardly change the results.
The employed values of the other parameters are summarized in Table \ref{parameter}.
\begin{table*}[htbp]
\begin{center}
\caption{\label{parameter}Standard values of the QCD parameters.
}
\begin{tabular}{cccc}
\hline\hline
 $\langle\bar{q}q\rangle$
&$m_0^2\equiv\langle\bar{q}g_s\vec{\sigma}\!\cdot\!\vec{G}q\rangle/\qq$
&$\langle\alpha_s\pi^{-1}G^2\rangle$
\\
\hline
 $(-0.23\;{\rm{GeV}})^3$ & $0.8{\rm{GeV}}^2$ & $(0.33 {\rm{GeV}})^4$\\
\hline\hline
\end{tabular}
\end{center}
\end{table*}

\begin{figure}[htbp]
\begin{center}
\includegraphics[width=20pc]{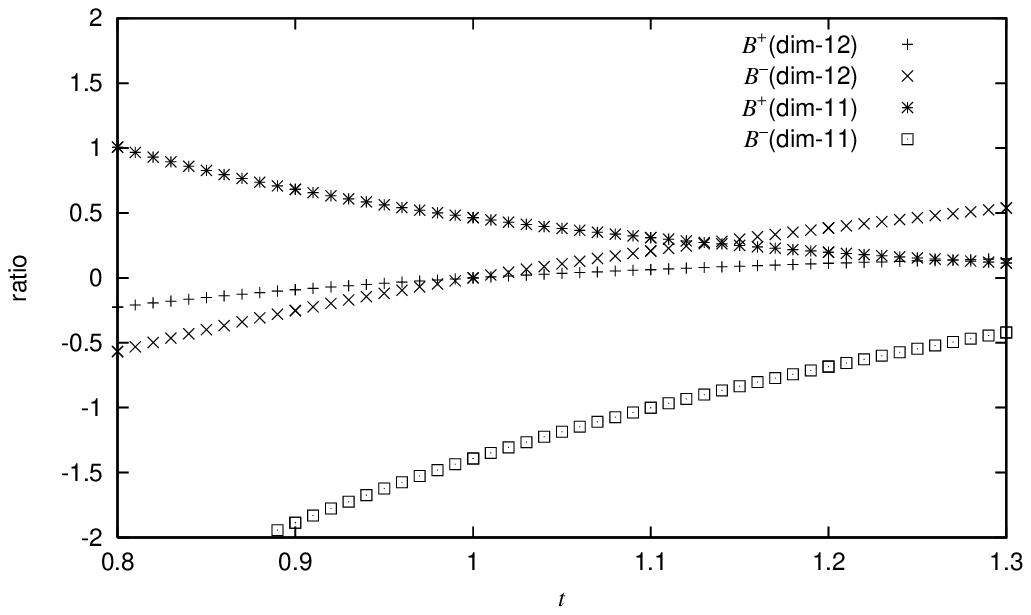}
\includegraphics[width=20pc]{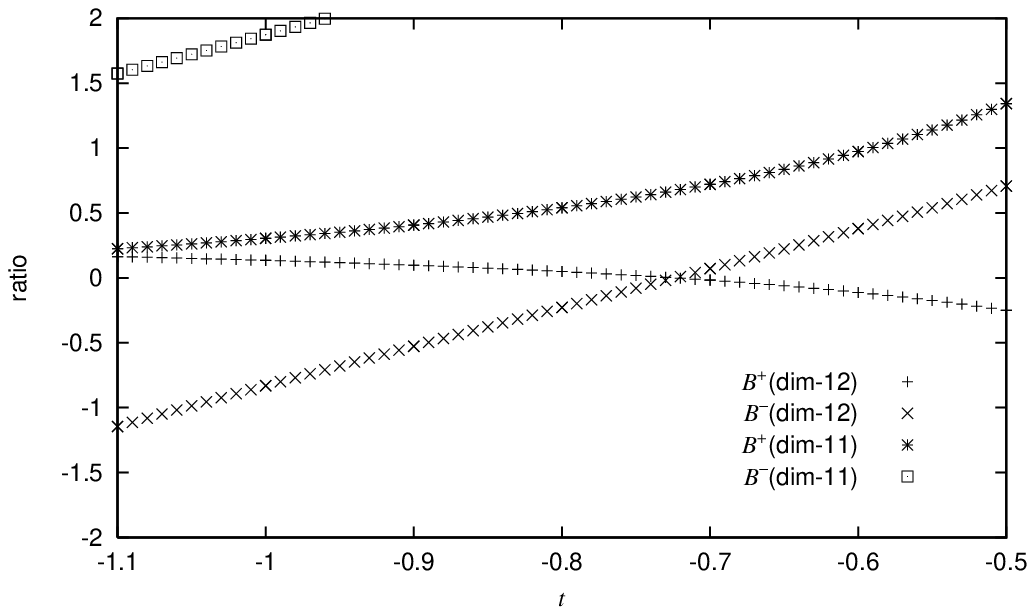}
\caption{Relative contributions of the dim.$-11$ and $ -12$ terms of the OPE 
as functions of $t$ with $M_B=1.2$ GeV and
$\sqrt{s_{{\rm{th}}}}=1.5$ GeV fixed. 
The left panel is for $t$ around unity and the right is for $t$ around $-1$.}\label{chi_re:1}
\end{center}
\end{figure}

The spectral functions are parametrized as Eq.~(\ref{SR_exo:11})  with a few physical parameters,  
the masses, $m_\pm$, and the residues, $|\lambda^\pm|^2$, of the designated pole states, 
and the threshold parameters,  $s^\pm_{\rm{th}}$.
The choice of the operator contains another parameter, $t$, defined in Eq.~(\ref{bpgeneral}).
We determine $t$ so that the operator couples to the positive- and negative-
parity states strongly and also that the contributions of higher-dimensional terms in the OPE are
small. 
Fig.~\ref{chi_re:1} shows the behavior of higher-dimensional terms normalized by the total OPE 
as a function of $t$.  $M_B=1.2\;{\rm{GeV}}$ and
$\sqrt{s_{{\rm{th}}}}=1,5\;{\rm{GeV}}$ are fixed. 
We find that around $t=1$ and $t=-1$, the higher-dimensional terms in the positive-parity nucleon sum rule
are suppressed. 
In view of the convergence of the OPE, such a $t$ is suitable for the sum rules. 
The three-quark part of the $t=-1$ operator is nothing but the ``Ioffe current'' operator, which is known to
work well in the nucleon sum rule (as well as the lattice QCD) calculations.

On the other hand, because the magnitude of the total OPE for the negative-parity nucleon
$B^-$ is smaller than that of $B^+$, 
the higher-dimensional terms give a relatively large contribution for the $B^-$ sum rule.
The region $t>1$ shows a fairly good convergence for the OPE of the negative parity nucleon.

%
%
%
\begin{figure}[htbp]
\begin{center}
\includegraphics[width=20pc]{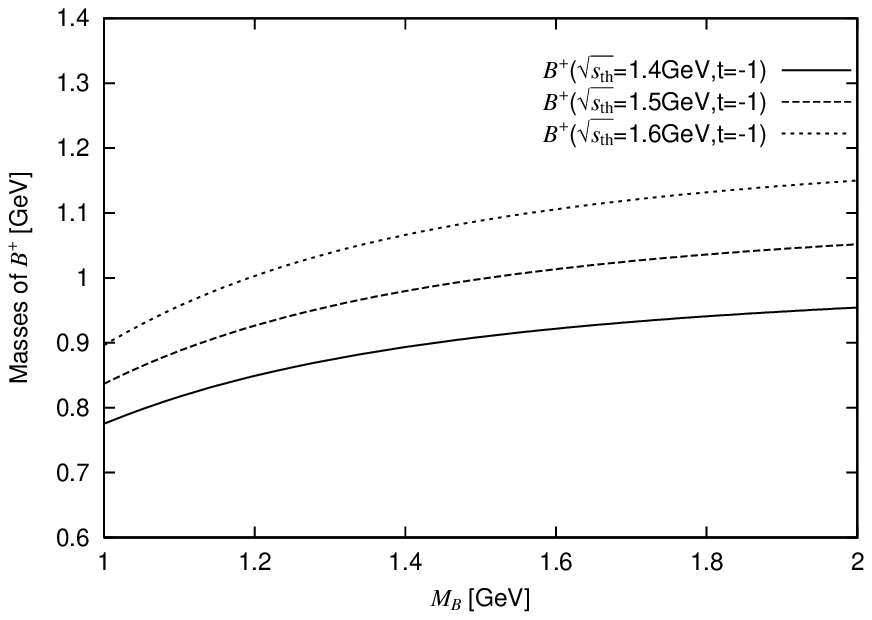}
\includegraphics[width=20pc]{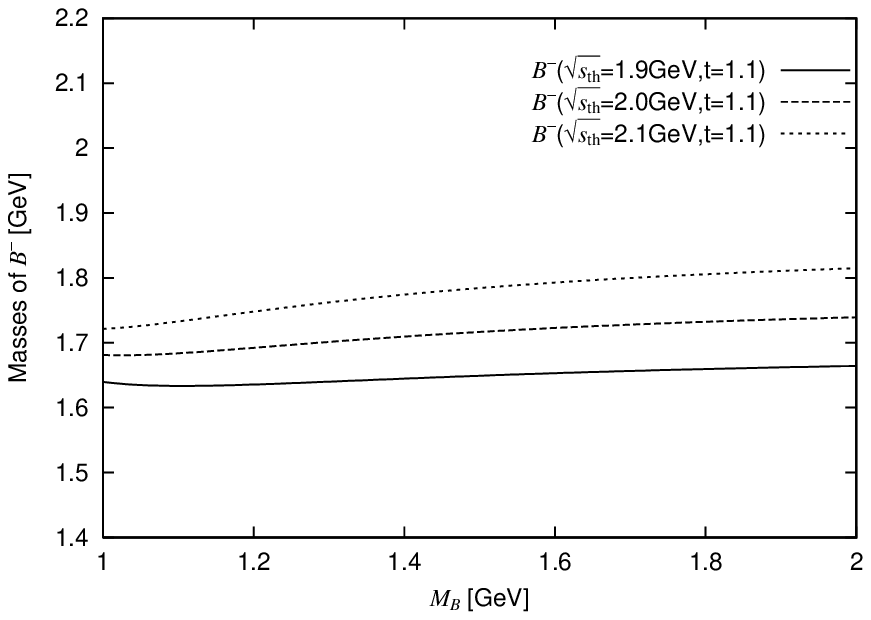}
\caption{The Borel mass dependences of the masses of $B^+$ and $B^-$.  
}\label{chi_re:3}
\end{center}
\end{figure}
\begin{figure}[htbp]
\begin{center}
\includegraphics[width=20pc]{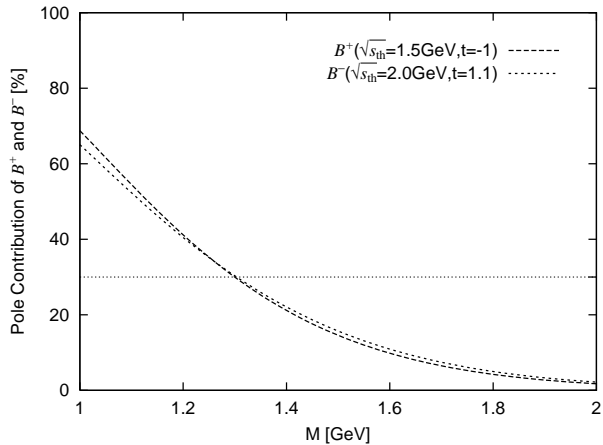}
\caption{The pole contributions defined by Eq. (\ref{chire:1}) for $B^+$ and $B^-$  as
functions of the Borel mass, $M_B$.}\label{chi_re:4}
\end{center}
\end{figure}

Another necessary check is to see whether the sum rule is consistent with the positive pole residues, $|\lambda_\pm|^2$.
Sometimes the sum rule gives spurious poles that do not correspond to physical states with positive pole residues.
The results may include some spurious solutions which yield negative residues.
In our case, we find that there is no physical solution at $t<0$ for $B^-$.

After the above considerations, we construct sum rules with $t=-1$ for positive-parity states and 
with $t=1.1$ for negative-parity states.
Then we determine the pole parameters as well as the threshold values, $\sqrt{s_{\rm th}}$, from the
sum rules under the condition that the Borel mass dependences are suppressed.
Fig. \ref{chi_re:3} shows the Borel mass dependences of the masses of $B^+$ and $B^-$. 
One sees that the Borel stability is fairly well satisfied.

In order to examine how well these sum rules work, we consider the pole contribution defined by
\begin{eqnarray}
\left(\int^{\sqrt{s_{{\rm{th}}}}}_0{\rm{d}}q_0\/\rho^\pm_{{\rm{OPE}}}(q_0)e^{-\frac{q_0^2}{M_B^2}}\right)
\Bigg/\left(\int^{\infty}_0{\rm{d}}q_0\/\rho^\pm_{{\rm{OPE}}}(q_0)e^{-\frac{q_0^2}{M_B^2}}\right) .
\label{chire:1}
\end{eqnarray}
Fig. \ref{chi_re:4} shows the ratios for the positive- and negative-parity sum rules.
We choose the Borel window as the region where the pole contribution is more than $30\%$. 
This constraint is somewhat weaker than the standard criterion used in the literature\cite{RRY1985}, 
but we have confirmed that the results are not sensitive to the choice of this value.
Then the chosen Borel window is at $M_B<1.3{\rm{GeV}}$.
The predicted masses of the $B^+$ and $B^-$ are about $0.8\sim 1.0{\rm{GeV}}$ and 
$1.6\sim 1.8{\rm{GeV}}$, respectively. 
The ambiguities are due to the Borel mass dependences and also from the ambiguities
in determining the threshold masses, which are not determined precisely because 
the Borel stability is not sensitive to the choice of the threshold, as is seen in Fig. \ref{chi_re:3}

\subsection{The sum rules for the ``pure" five-quark operator}

Although the above results seem to indicate the existence of the  ``exotic" nucleon,
it is noted that the predicted mass of $B^+$ is very low.
We suspect that it corresponds to the nucleon ground state ($N(939)$), 
which is known to be reproduced by a three-quark operator very well. 
Indeed, the interpolating field $B^\ast$ happens to contain a component which is
equivalent to the standard three-quark operator after contracting the $\langle\bar{q}q\rangle$ condensate.
Then, the above results may just reproduce those obtained from the three-quark operator.

In order to investigate the above possibility, we separate the three-quark operator by making  
all the possible contractions of $\bar q q$ and obtain
\begin{eqnarray}
B^\ast(x)=B^{s\ast}(x)+\underbrace{2\langle\bar{q}q\rangle B(x)}_{B^3(x)}.
\label{chire:2}
\end{eqnarray}
Here, the interpolating field $B(x)$ is the standard three-quark operator, Eq.~(\ref{bpgeneral}).
Then we expect that the interpolating field $B^{s\ast}(x)$ is a genuine five-quark operator in which the three-quark
component is subtracted.

The interpolating field $B(x)$ has been employed in standard QCD sum rule analyses of 
positive- and negative-parity nucleon states \cite{Ioffe,Jido:1996ia}.
(The operator $B^3(x)$, of course, gives the identical sum rule.)
We indeed find that the $B(x)$ (or $B^3(x)$) sum rule gives results quite similar to those obtained
from the $B^\ast$ sum rule.  
Namely, the above results are nothing but 
the reflection of the contaminating $B^3(x)$ operator in the five-quark operator.
We thus conclude that the obtained masses of the positive- and negative-parity nucleons are 
those of the states with standard chiral property, not a candidates for the ``exotic'' one.

Expecting that the interpolating field $B^{s\ast}$ as the pure five-quark
component should couple strongly to the ``exotic'' state, we perform the sum rule
analysis of $B^{s\ast}$.
From Eqs. (\ref{SR_exo:10}) and (\ref{chire:2}),
the OPE side of the sum rule for $B^{s\ast}$ is given by
\begin{eqnarray}
A^{s\ast}_{{\rm{OPE}}}(q_0)&=&
+(90+36t+90t^2)\frac{q_0^{11}}{2^{11}\times 5!\;7!\pi^8}
\nonumber \\
&&+(41+14t+41t^2)\frac{q_0^7}{2^{10}\times 4!\;5!\pi^6}\langle\alpha_s\pi^{-1}G^2\rangle
\nonumber \\
&&+(-3-2t+11t^2)\frac{q_0^5}{2^{12}\pi^4}\langle\bar{q}q\rangle^2
\nonumber \\
&&-(-25+38t+203t^2)\frac{q_0^3}{3\!\times\!2^{12}\pi^4}\langle\bar{q}q\rangle\langle\bar{q}g_s
\vec{\sigma}\!\cdot\!\vec{G}q\rangle
\nonumber \\
&&+(-197-122t+499t^2)\frac{q_0}{3^2\!\times\!2^{12}\pi^2}\langle\bar{q}q\rangle^2\langle\alpha_s\pi^{-1}G^2\rangle
\nonumber \\
&&+(-3539+1150t+6853t^2)\frac{q_0}{3^2\!\times\!2^{18}\pi^4}\langle\bar{q}g_s\vec{\sigma}\!\cdot\!\vec{G}
q\rangle^2
\nonumber \\
&&+(-11-2t+13t^2)\frac{q_0}{3^2\!\times\!2^4}\langle\bar{q}q\rangle^4\delta(q^2),
\nonumber \\  
B^{s\ast}_{{\rm{OPE}}}(q_0)&=&
+(-33-6t+39t^2)\frac{q_0^8}{2^{10}\times 4!\;5!\pi^6}\langle\bar{q}q\rangle 
\nonumber \\
&&-(-13-7t+20t^2)\frac{q_0^6}{2^{15}\times 3^2\pi^6}\langle\bar{q}g_s\vec{\sigma}\!\cdot\!\vec{G}q\rangle
\nonumber \\
&&+(-28+5t+23t^2)\frac{q_0^4}{3^2\!\times\!2^{12}\pi^4}\langle\bar{q}q\rangle\langle\alpha_s\pi^{-1}G^2\rangle
\nonumber \\
&&+(-3-2t+11t^2)\frac{q_0^2}{2^7\pi^2}\langle\bar{q}q\rangle^3
\nonumber \\
&&+(-1+14t-169t^2)\frac{1}{3\!\times\!2^9\pi^2}\langle\bar{q}q\rangle^2\langle\bar{q}g_s\vec{\sigma}
\!\cdot\!\vec{G}q\rangle.
\label{chire:4}
\end{eqnarray}

\begin{figure}[htbp]
\begin{center}
\includegraphics[width=20pc]{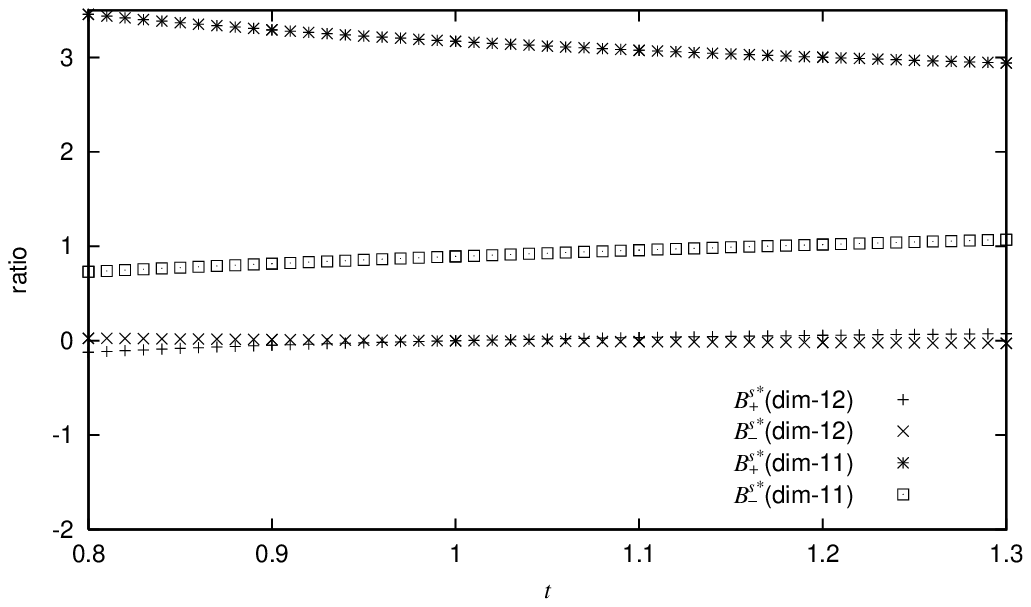}
\includegraphics[width=20pc]{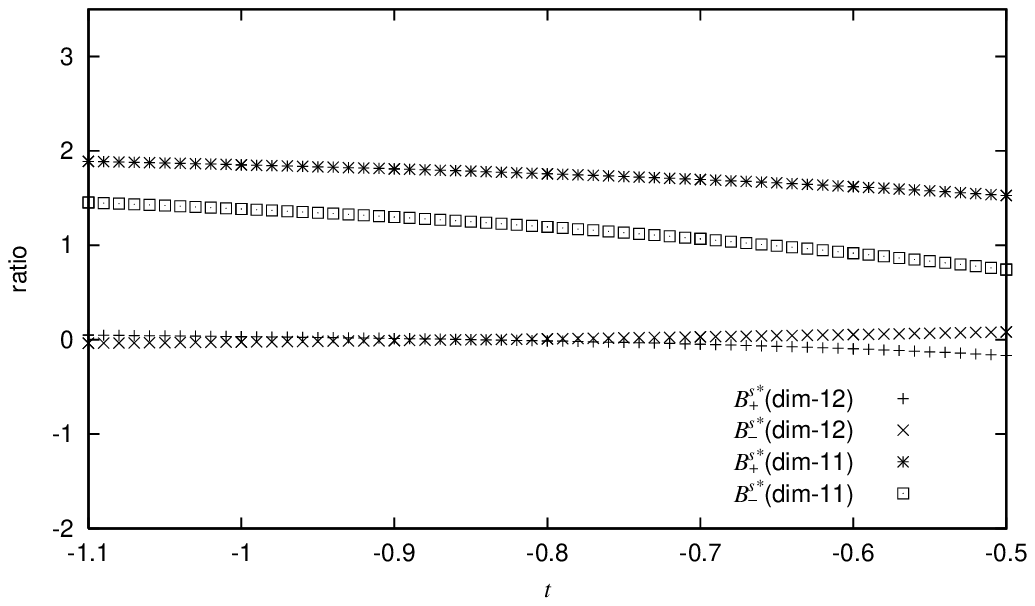}
\caption{Relative contributions of the dim.$-11$ and $-12$ terms of the OPE for $B^{s*}$
as functions of $t$ with $M_B=1.2$ GeV and
$\sqrt{s_{{\rm{th}}}}=1.5$ GeV fixed. 
The left panel is for $t$ around unity and the right is for $t$ around $-1$.}\label{chi_re:9}
\end{center}
\end{figure}
%
%
%
%
\begin{figure}[htbp]
\begin{center}
\includegraphics[width=20pc]{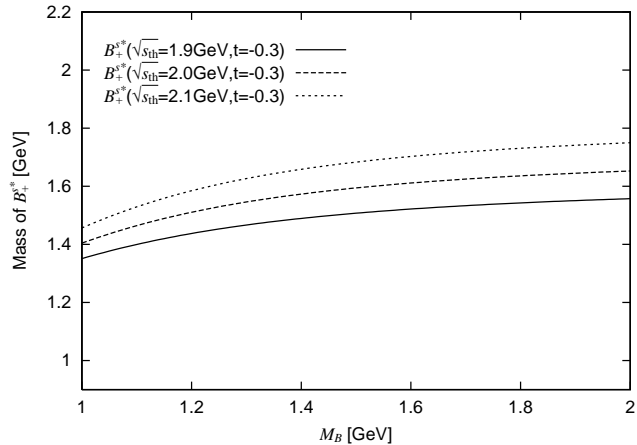}
\caption{The Borel mass dependences of the masses of $B^{s*}_+$ for various $\sqrt{s_{\rm{th}}}$.}\label{chi_re:11}
\end{center}
\end{figure}
\begin{figure}[htbp]
\begin{center}
\includegraphics[width=20pc]{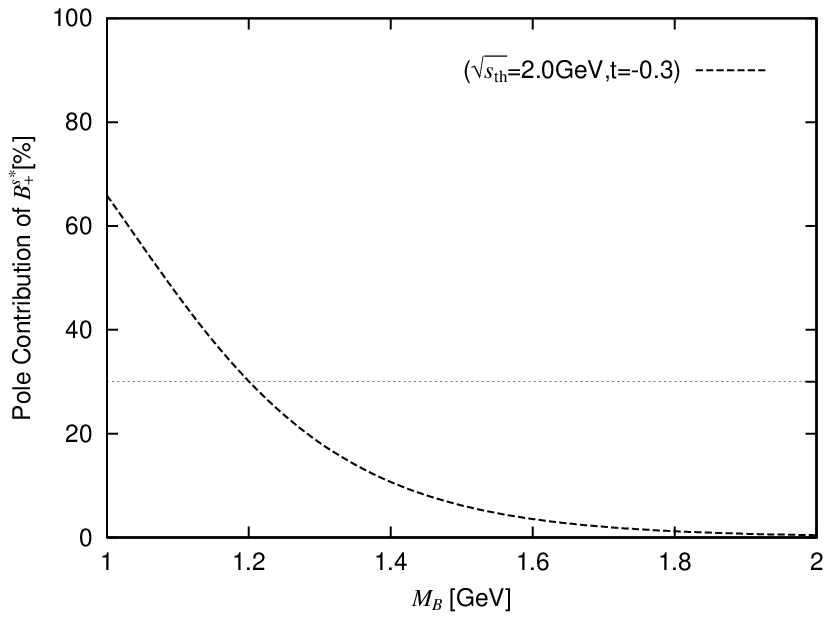}
\caption{The pole contribution defined by Eq. (\ref{chire:1}) for $B^{s\ast}_+$  as
a function of the Borel mass, $M_B$.}
\label{chi_re:12}
\end{center}
\end{figure}

In Fig. \ref{chi_re:9}, we show the convergence of higher-dimensional contributions for $B^{s\ast}_+$
and $B^{s\ast}_-$. We find no good OPE convergence in the region $-2<t<2$ due to cancellation
among the OPE terms. In particular, the dimension 11 contribution is larger than the total OPE.
This bad convergence makes the sum rule results less valid.
All the solutions for the negative-parity nucleon $B^{s\ast}_-$ are 
found to be spurious because the pole residue is negative. 
Therefore we conclude that the interpolating field $B^{s\ast}(x)$ does not support the picture that 
the negative-parity nucleon $B^{s\ast}_-$ transforms under the ``exotic" chiral transformation rule.

For the positive-parity nucleon $B^{s\ast}_+$, we find a physical solution, which has 
a positive residue at around $t=-0.3$. 
In Figs. \ref{chi_re:11} and \ref{chi_re:12},
the masses and the pole contribution for the positive-parity nucleon $B^{s\ast}_+$ are plotted,
respectively.
A fairly good Borel stability is observed, and with the predicted mass, about $1.4\sim 1.6{\rm{GeV}}$.
However, because of the bad convergence of the OPE and the small pole contribution, we have
to conclude that the validity for the solution of the positive-parity nucleon $B^{s\ast}_+$ is poor
and that the sum rule cannot confirm the five-quark exotic nucleon state.

\section{Conclusion}\label{chiral_sum}

We have studied the nucleon sum rules using the five-quark operator which is transformed 
in an ``exotic'' way under the chiral transformation.  The operator has been shown to have
negative axial charge, $g_A =-1$, and thus may couple to a nucleon resonance which forms
a chiral-mirror pair with a standard nucleon state.

Using the five-quark interpolating field operator, $B^\ast(x)$, defined in Eq. (\ref{eq:Bmirror}),
we have constructed the sum rule.  
We have employed the parity-projected sum rule approach and examined the positive- and negative-parity 
nucleon resonances separately.
After checking the reliability of the sum rules, we have obtained
the signal of the positive-parity nucleon at the mass around $0.8\sim 1.0{\rm{GeV}}$.
The negative-parity nucleon is also obtained at the mass around $1.6\sim 1.8{\rm{GeV}}$.
It is, however, argued that these signals come not from the five-quark nucleon states, but the ordinary
three-quark nucleons due to the contamination of three-quark operators in $B^\ast(x)$.

We then have extracted the genuine five-quark operator $B^{s\ast}(x)$ by subtracting the three-quark operator 
multiplied by the $\langle\bar{q}q\rangle$ condensate.
The sum rules for the positive- and negative-parity nucleons are constructed from $B^{s\ast}(x)$, 
which may couple  strongly to the ``exotic" nucleons.
No negative-parity nucleon state is found  in the energy region below 2 GeV, which can be
concluded from the fact that the analysis results in negative pole residues.
A possible physical solution of the positive-parity nucleon state is obtained for the operator 
with $t=-0.3$, where $t$ is the parameter in the interpolating field operator representing the
mixing of two possible nucleon operators. 
The predicted mass is about $1.4\sim 1.6
{\rm{GeV}}$ but the pole contribution and the convergence of the OPE are not sufficient 
to conclude that the positive parity ``exotic'' nucleon exists.

If the positive-parity state $B^{s\ast}_+$ observed for the interpolating field $B^{s\ast}$
is a physical state, it is possible that  this state is assigned to the first positive-parity 
nucleon resonance, $N(1440)$, and that $N(1440)$ is the state
with the ``exotic" chiral property.
In such a case, there are, at least, two candidates for its chiral-mirror partner,
i.e.,  two $1/2^-$ states, $N(1535)$ and $N(1650)$.
This conjecture is consistent with the recent claim that both of them are considered to 
be three-quark-like nucleon resonances\cite{Hyodo}.

Unfortunately, as is mentioned above, the validity of our sum rule conclusion is not high.
One possibility to improve this situation is to use other choices of the interpolating field
operator.
In the present case, the choice of the five-quark operator with ``exotic'' chiral property is not
unique, but there are several independent operators which satisfy the commutation
relation Eq.(\ref{ga-1}).
Further study is desired using other operators or possible linear combinations.

\acknowledgments
This work was partially supported by KAKENHI, 17070002 (Priority area) and 19540275.
We thank Drs. Daisuke Jido, Atsushi Hosaka and Jun Sugiyama for discussions.
T. N. acknowledges the support from the 21st Century COE Program 
at TokyoTech.~``Nanometer-Scale Quantum Physics'' by the MEXT, Japan.
The support for P.G. from the Ito Foundation for International
Education Exchange is greatfully acknowledged.

\end{document}